\documentclass{article}
\usepackage{spconf,amsmath,amsfonts,graphicx}
\usepackage{amssymb,textcomp,mathtools}
\usepackage{bm,upgreek,algorithm,hyperref}
\usepackage{multirow,booktabs,hhline,array}
\usepackage{cite,url,makecell,setspace}

\pdfoutput=1

\def\thline{\noalign{\hrule height 1.0pt}}

\renewcommand{\vec}[1]{\bm{\mathrm{#1}}}

\title{Dual-path RNN: efficient long sequence modeling for\\time-domain single-channel speech separation}

\name{Yi Luo$^{\dag}$\sthanks{Work done during internship at Microsoft Research.}, Zhuo Chen$^\ddagger$, Takuya Yoshioka$^\ddagger$}
\address{$^\dag$Department of Electrical Engineering, Columbia University, NY, USA\\$^\ddagger$Microsoft, One Microsoft Way, Redmond, WA, USA}

\begin{document}
\ninept
\maketitle

\begin{abstract}
Recent studies in deep learning-based speech separation have proven the superiority of time-domain approaches to conventional time-frequency-based methods. Unlike the time-frequency domain approaches, the time-domain separation systems often receive input sequences consisting of a huge number of time steps, which introduces challenges for modeling extremely long sequences. Conventional recurrent neural networks (RNNs) are not effective for modeling such long sequences due to optimization difficulties, while one-dimensional convolutional neural networks (1-D CNNs) cannot perform utterance-level sequence modeling when its receptive field is smaller than the sequence length. In this paper, we propose dual-path recurrent neural network (DPRNN), a simple yet effective method for organizing RNN layers in a deep structure to model extremely long sequences. DPRNN splits the long sequential input into smaller chunks and applies intra- and inter-chunk operations iteratively, where the input length can be made proportional to the square root of the original sequence length in each operation. Experiments show that by replacing 1-D CNN with DPRNN and apply sample-level modeling in the time-domain audio separation network (TasNet), a new state-of-the-art performance on WSJ0-2mix is achieved with a 20 times smaller model than the previous best system.
\end{abstract}


\section{Introduction}
\label{sec:intro}
Recent progress in deep learning-based speech separation has ignited the interest of the research community in time-domain approaches \cite{luo2018tasnet, stoller2018wave, venkataramani2018end, luo2019conv, shi2019furcanext, bahmaninezhad2019comprehensive}. Compared with standard time-frequency domain methods, time-domain methods are designed to jointly model the magnitude and phase information and allow direct optimization with respect to both time- and frequency-domain differentiable criteria \cite{fu2018end, le2019sdr, luo2019fasnet}.

Current time-domain separation systems can be mainly categorized into \textit{adaptive front-end} and \textit{direct regression} approaches. The \textit{adaptive front-end} approaches aim at replacing the short-time Fourier transform (STFT) with a differentiable transform to build a front-end that can be learned jointly with the separation network. Separation is applied to the front-end output as with the conventional time-frequency domain methods applying the separation processes to spectrogram inputs \cite{venkataramani2018end, luo2019conv, shi2019furcanext}. Being independent of the traditional time-frequency analysis paradigm, these systems are able to have a much more flexible choice on the window size and the number of basis functions for the front-end. On the other hand, the \textit{direct regression} approaches learn a regression function from an input mixture to the underlying clean signals without an explicit front-end, typically by using some form of one-dimensional convolutional neural networks (1-D CNNs) \cite{stoller2018wave, lluis2018end, fu2018end}.

A commonality between the two categories is that they both rely on effective modeling of
extremely long input sequences. The direct regression methods perform separation at the waveform sample level, while the number of the samples can usually be tens of thousands, or sometimes even more. The performance of the adaptive front-end methods also depend on selection of the window size, where a smaller window improves the separation performance at the cost of a significantly longer front-end representation \cite{luo2019conv, kavalerov2019universal}. This poses an additional challenge as conventional sequential modeling networks, including RNNs and 1-D CNNs,  have difficulty on learning such long-term temporal dependency \cite{li2018independently}. Moreover, unlike RNNs that have dynamic receptive fields, 1-D CNNs with fixed receptive fields that are smaller than the sequence length are not able to fully utilize the sequence-level dependency \cite{bai2018empirical}.

In this paper, we propose a simple network architecture, which we refer to as \textit{dual-path RNN (DPRNN)}, that organizes any kinds of RNN layers to model long sequential inputs in a very simple way. The intuition is to split the input sequence into shorter chunks and interleave two RNNs, an \textit{intra-chunk} RNN and an \textit{inter-chunk} RNN, for local and global modeling, respectively. In a DPRNN block, the intra-chunk RNN first processes the local chunks independently, and then the inter-chunk RNN aggregates the information from all the chunks to perform utterance-level processing. For a sequential input of length $L$, DPRNN with chunk size $K$ and chunk hop size $P$ contains $S$ chunks, where $K$ and $S$ corresponds to the input lengths for the inter- and intra-chunk RNNs, respectively. When $K \approx S$, the two RNNs have a sublinear input length ($O(\sqrt L)$) as opposed to the original input length ($O(L)$), which greatly decreases the optimization difficulty that arises when $L$ is extremely large.

Compared with other approaches for arranging local and global RNN layers, or more general the hierarchical RNNs that perform sequence modeling in multiple time scales \cite{li2015hierarchical, chung2016hierarchical, mehri2016samplernn, ishiwatari2017chunk, zhou2017chunk, chang2017dilated}, the stacked DPRNN blocks \textit{iteratively and alternately} perform the intra- and inter-chunk operations, which can be treated as an interleaved processing between local and global inputs. Moreover, the first RNN layer in most hierarchical RNNs still receives the entire input sequence, while in stacked DPRNN each intra- or inter-chunk RNN receives the same sublinear input size across all blocks. Compared with CNN-based architectures such as temporal convolutional networks (TCNs) that only perform local modeling due to the fixed receptive fields \cite{luo2019conv, shi2019furcanext, liu2019divide}, DPRNN is able to fully utilize global information via the inter-chunk RNNs and achieve superior performance with an even smaller model size. In Section~\ref{sec:results} we will show that by simply replacing TCN by DPRNN in a previously proposed time-domain separation system \cite{luo2019conv}, the model is able to achieve a 0.7 dB (4.6\%) relative improvement with respect to scale-invariant signal-to-noise ratio (SI-SNR)  \cite{le2019sdr} on WSJ0-2mix with a 49\% smaller model size. By performing the separation at the waveform sample level, i.e. with window size of 2 samples and hop size of 1 sample, a new state-of-the-art performance is achieved with a 20 times smaller model than the previous best system.

\section{Dual-path Recurrent Neural Network}
\label{sec:model}
\begin{figure*}[ht]
	\small
	\centering
	\includegraphics[width=1.8\columnwidth]{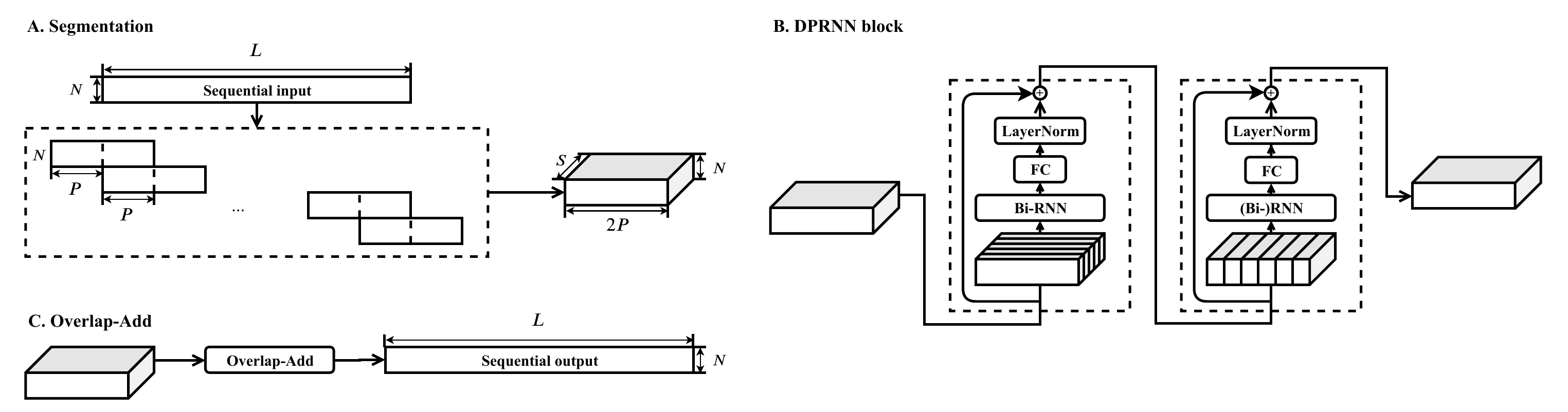}
	\caption{System flowchart of dual-path RNN (DPRNN). (A) The segmentation stage splits a sequential input into chunks with or without overlaps and concatenates them to form a 3-D tensor. In our implementation, the overlap ratio is set to 50\%. (B) Each DPRNN block consists of two RNNs that have recurrent connections in different dimensions. The \textit{intra-chunk} bi-directional RNN is first applied to individual chunks in parallel to process local information. The \textit{inter-chunk} RNN is then applied across the chunks to capture global dependency. Multiple blocks can be stacked to increase the total depth of the network. (C) The 3-D output of the last DPRNN block is converted back to a sequential output by performing overlap-add on the chunks.}
	\label{fig:flowchart}
\end{figure*}

\subsection{Model design}

A dual-path RNN (DPRNN) consists of three stages: \textit{segmentation}, \textit{block processing}, and \textit{overlap-add}. The segmentation stage splits a sequential input into overlapped chunks and concatenates all the chunks into a 3-D tensor. The tensor is then passed to stacked DPRNN blocks to iteratively apply local (intra-chunk) and global (inter-chunk) modeling in an alternate fashion. The output from the last layer is transformed back to a sequential output with overlap-add method. Figure~\ref{fig:flowchart} shows the flowchart of the model.

\subsubsection{Segmentation}

For a sequential input $\vec{W} \in \mathbb{R}^{N\times L}$ where $N$ is the feature dimension and $L$ is the number of time steps, the segmentation stage splits $\vec{W}$ into chunks of length $K$ and hop size $P$. The first and last chunks are zero-padded so that every sample in $\vec{W}$ appears and only appears in $K/P$ chunks, generating $S$ equal size chunks $\vec{D}_s \in \mathbb{R}^{N\times K}, \, s = 1, \ldots, S$. All chunks are then concatenated together to form a 3-D tensor $\vec{T} = [\vec{D}_1, \ldots, \vec{D}_{S}] \in \mathbb{R}^{N\times K\times S}$.

\subsubsection{Block processing}

The segmentation output $\vec{T}$ is then passed to the stack of $B$ DPRNN blocks. 
Each block transforms an input 3-D tensor into another tensor with the same shape. 
We denote the input tensor for block $b = 1, \ldots, B$ as $\vec{T}_b \in \mathbb{R}^{N\times K\times S}$, where $\vec{T}_1 = \vec{T}$. Each block contains two sub-modules corresponding to intra- and inter-chunk processing, respectively. The intra-chunk RNN is always bi-directional and is applied to the second dimension of $\vec{T}_b$, i.e., within each of the $S$ blocks:
\begin{align}
	\vec{U}_b = [f_b(\vec{T}_b[:,:,i]), \, i=1, \ldots, S]
\end{align}
where $\vec{U}_b \in \mathbb{R}^{H\times K\times S}$ is the output of the RNN, $f_b(\cdot)$ is the mapping function defined by the RNN, and $\vec{T}_b[:,:,i] \in \mathbb{R}^{N\times K}$ is the sequence defined by chunk $i$. A linear fully-connected (FC) layer is then applied to transform the feature dimension of $\vec{U}_b$ back to that of $\vec{T}_b$
\begin{align}
	\hat{\vec{U}}_b = [\vec{G}\vec{U}_b[:,:,i] + \vec{m}, \, i=1, \ldots, S]
\end{align}
where $\hat{\vec{U}} \in \mathbb{R}^{N\times K\times S}$ is the transformed feature, $\vec{G} \in \mathbb{R}^{N\times H}$ and $\vec{m} \in \mathbb{R}^{N\times 1}$ are the weight and bias of the FC layer, respectively, and $\vec{U}_b[:,:,i] \in \mathbb{R}^{H\times K}$ represents chunk $i$ in $\vec{U}_b$. Layer normalization (LN) \cite{ba2016layer} is then applied to $\hat{\vec{U}}$, which we empirically found to be important for the model to have a good generalization ability:
\begin{align}
	LN(\hat{\vec{U}}_b) &= \frac{\hat{\vec{U}}_b - \mu(\hat{\vec{U}}_b)}{\sqrt{\sigma(\hat{\vec{U}}_b)+\epsilon}} \odot \vec{z} + \vec{r} \\
\end{align}
where $\vec{z}, \vec{r} \in \mathbb{R}^{N\times 1}$ are the rescaling factors, $\epsilon$ is a small positive number for numerical stability, and $\odot$ denotes the Hadamard product. $\mu(\cdot)$ and $\sigma(\cdot)$ are the mean and variance of the 3-D tensor defined as
\begin{align}
    \mu(\hat{\vec{U}}_b) &= \frac{1}{NKS}\sum_{i=1}^{N}\sum_{j=1}^{K}\sum_{s=1}^{S} \hat{\vec{U}}_b[i,j,s] \\
	\sigma(\hat{\vec{U}}_b) &= \frac{1}{NKS}\sum_{i=1}^{N}\sum_{j=1}^{K}\sum_{s=1}^{S} (\hat{\vec{U}}_b[i,j,s] - \mu(\hat{\vec{U}}_b))^2
\end{align}

A residual connection is then added between the output of LN operation and the input $\vec{T}_b$:
\begin{align}
	\hat{\vec{T}}_b = \vec{T}_b + LN(\hat{\vec{U}}_b)
\end{align}
$\hat{\vec{T}}_b$ is then served as the input to the inter-chunk RNN sub-module, where the RNN is applied to the last dimension, i.e. the aligned $K$ time steps in each of the $S$ blocks:
\begin{align}
\vec{V}_b = [h_b(\hat{\vec{T}}_b[:,i,:]), \, i=1, \ldots, K]
\end{align}
where $\vec{V}_b \in \mathbb{R}^{H\times K\times S}$ is the output of RNN, $h_b(\cdot)$ is the mapping function defined by the RNN, and $\hat{\vec{T}}_b[:,i,:] \in \mathbb{R}^{N\times S}$ is the sequence defined by the $i$-th time step in all $S$ chunks. As the intra-chunk RNN is bi-directional, each time step in $\hat{\vec{T}}_b$ contains the entire information of the chunk it belongs to, which allows the inter-chunk RNN to perform fully sequence-level modeling. As with the intra-chunk RNN, a linear FC layer and the LN operation are applied on top of $\vec{V}_b$. A residual connection is also added between the output and $\hat{\vec{T}}_b$ to form the output for DPRNN block $b$. For $b < B$, the output is served as the input to the next block $\vec{T}_{b+1}$.

\subsubsection{Overlap-Add}

Denote the output of the last DPRNN block as $\vec{T}_{B+1} \in \mathbb{R}^{N\times K \times S}$. To transform it back to a sequence, the overlap-add method is applied to the $S$ chunks to form output $\vec{Q} \in \mathbb{R}^{N\times L}$.

\subsection{Discussion}
\label{sec:dprnn_dis}

Consider the sum of the input sequence lengths for the intra- and inter-chunk RNNs in a single block denoted by $K+S$ where the hop size is set to be 50\% (i.e. $P=K/2$) as in Figure~\ref{fig:flowchart}. It is simple to see that $S = \lceil2L/K\rceil+1$ where $\lceil\cdot\rceil$ is the ceiling function. To achieve minimum total input length $K+S=K+\lceil2L/K\rceil+1$, $K$ should be selected such that $K\approx \sqrt{2L}$, and then $S$ also satisfies $S\approx \sqrt{2L} \approx K$. This gives us sublinear input length ($O(\sqrt L)$) rather than the original linear input length ($O(L)$).

For tasks that require online processing, the inter-chunk RNN can be made uni-directional, scanning from the first up to the current chunks. The later chunks can still utilize the information from all previous chunks, and the minimal system latency is thus defined by the chunk size $K$. This is unlike standard CNN-based models that can only perform local processing due to the fixed receptive field or conventional RNN-based models that perform frame-level instead of chunk-level modeling. The performance difference between the online and offline settings, however, is beyond the scope of this paper.

\section{Experimental procedures}
\label{sec:exp}
\subsection{Model configurations}

Although DPRNN can be applied to any systems that require long-term sequential modeling, we investigate its application to the time-domain audio separation network (TasNet) \cite{luo2018tasnet, luo2018real, luo2019conv}, an adaptive front-end method that achieves high speech separation performance on a benchmarking dataset. TasNet contains three parts: (1) a linear 1-D convolutional encoder that encapsulates the input mixture waveform into an adaptive 2-D front-end representation, (2) a separator that estimates $C$ masking matrices for $C$ target sources, and (3) a linear 1-D transposed convolutional decoder that converts the masked 2-D representations back to waveforms. We use the same encoder and decoder design as in \cite{luo2019conv} while the number of filters is set to be 64. As for the separator, we compare the proposed deep DPRNN with the optimally configured TCN described in \cite{luo2019conv}. We use 6 DPRNN blocks using BLSTM \cite{hochreiter1997long} as the intra- and inter-chunk RNNs with 128 hidden units in each direction. The chunk size $K$ for DPRNN is defined empirically according to the length of the front-end representation such that $K\approx \sqrt{2L}$ in the training set as discussed in Section~\ref{sec:dprnn_dis}.

\subsection{Dataset}

We evaluate our approach on two-speaker speech separation and recognition tasks. The separation-only experiment is conducted on the widely-used WSJ0-2mix dataset \cite{hershey2016deep}. WSJ0-2mix contains 30 hours of 8k Hz training data that are generated from the Wall Street Journal (WSJ0) si\_tr\_s set. It also has 10 hours of validation data and 5 hours of test data generated by using the si\_dt\_05 and si\_et\_05 sets, respectively. Each mixture is artificially generated by randomly selecting different speakers from the corresponding set and mixing them at a random relative signal-to-noise ratio (SNR) between -5 and 5 dB. For the speech separation and recognition experiment, we create 200 hours and 10 hours of artificially mixed noisy reverberant mixtures sampled from the Librispeech dataset \cite{panayotov2015librispeech} for training and validation, respectively. The 16 kHz signals are convolved with room impulse responses generated by the image method \cite{allen1979image}. The length and width of the room are randomly sampled between 2 and 10 meters, and the height is randomly sampled between 2 and 5 meters. The reverberation time (T60) is randomly sampled between 0.1 and 0.5 seconds. The locations for the speakers as well as the single microphone are all randomly sampled. The two reverberated signals are rescaled to a random SNR between -5 and 5 dB, and further shifted such that the overlap ratio between the two speakers is 50\% on average. The resultant mixture is further corrupted by random isotropic noise at a random SNR between 10 and 20 dB \cite{habets2007generating}. For evaluation, we generate mixture in the same manner sampled from Microsoft's internal gender-balanced clean speech collection consisting of 44 speakers. The target for separation is the reverberant clean speech for both speakers.

\subsection{Experiment configurations}

We train all models for 100 epochs on 4-second long segments. The learning rate is initialized to $1e^{-3}$ and decays by 0.98 for every two epochs. Early stopping is applied if no best model is found in the validation set for 10 consecutive epochs. Adam \cite{kingma2014adam} is used as the optimizer. Gradient clipping with maximum $L_2$-norm of 5 is applied for all experiments. All models are trained with utterance-level permutation invariant training (uPIT) \cite{kolbaek2017multitalker} to maximize scale-invariant SNR (SI-SNR) \cite{le2019sdr}.

The effectiveness of the systems is assessed both in terms of signal fidelity and speech recognition accuracy. The degree of improvement in the signal fidelity is measured by signal-to-distortion ratio improvement (SDRi) \cite{vincent2006performance} as well as SI-SNR improvement (SI-SNRi). The speech recognition accuracy is measured by the word error rate (WER) on both separated speakers.

\section{Results and discussions}
\label{sec:results}
\subsection{Results on WSJ0-2mix}

We first report the results on the WSJ0-2mix dataset. Table~\ref{tab:wsj0-tasnet} compares the TasNet-based systems with different separator networks. We can see that simply replacing TCN by DPRNN improves the separation performance by 4.6\% with a 49\% smaller model. This shows the superiority of the proposed local-global modeling to the previous CNN-based local-only modeling. Moreover, the performance can be consistently improved by further decreasing the filter length (and the hop size as a consequence) in the encoder and decoder. The best performance is obtained when the filter length is 2 samples with an encoder output of more than 30000 frames. This can be extremely hard or even impossible for standard RNNs or CNNs to model, while with the proposed DPRNN the use of such a short filter becomes possible and achieves the best performance.

Table~\ref{tab:wsj0-compare} compares the DPRNN-TasNet with other previous systems on WSJ0-2mix. We can see that DPRNN-TasNet achieves a new record on SI-SNRi with a 20 times smaller model than FurcaNeXt \cite{shi2019furcanext}, the previous state-of-the-art system. The small model size and the superior performance of DPRNN-TasNet indicate that speech separation on WSJ0-2mix dataset can be solved without using enormous or complex models, revealing the need for using more challenging and realistic datasets in future research.

\begin{table}[!htbp]
	\small
	\centering
	\caption{Comparison of different separator networks and configurations on WSJ0-2mix in TasNet-based speech separation. Prior work used TCN-TasNet.}
	\label{tab:wsj0-tasnet}
	\begin{tabular}{c|c|c|c|c|c}
		\thline
		\thead{Separator \\ network} & \thead{Model \\ size} & \thead{Window \\ (samples)} & \thead{Chunk size \\ (frames)} & \thead{SI-SNRi \\ (dB)} & \thead{SDRi \\ (dB)} \\
		\hline
        TCN & 5.1M & 16 & -- & 15.2 & 15.5 \\
        \hline
        \multirow{4}{*}{DPRNN} & \multirow{4}{*}{\bf{2.6M}} & 16 & 100 & 16.0 & 16.2 \\
        & & 8 & 150 & 17.0 & 17.3 \\
        & & 4 & 200 & 17.9 & 18.1 \\
        & & 2 & 250 & \bf{18.8} & \bf{19.0} \\
		\thline
	\end{tabular}
\end{table}

\begin{table}[!htbp]
	\small
	\centering
	\caption{Comparison with other methods on WSJ0-2mix.}
	\label{tab:wsj0-compare}
	\begin{tabular}{c|c|c|c}
		\thline
		\thead{Method} & \thead{Model \\ size} & \thead{SI-SNRi\\ (dB)} & \thead{SDRi\\ (dB)} \\
		\thline
		DPCL++ \cite{isik2016single} & 13.6M & 10.8 & -- \\
		uPIT-BLSTM-ST \cite{kolbaek2017multitalker} & 92.7M &  -- & 10.0 \\
		ADANet \cite{luo2017speaker} & 9.1M & 10.4 & 10.8 \\
		WA-MISI-5 \cite{wang2018end} & 32.9M & 12.6 & 13.1 \\
		Conv-TasNet-gLN \cite{luo2019conv} & 5.1M & 15.3 & 15.6 \\
		Sign Prediction Net \cite{wang2019deep} & 55.2M & 15.3 & 15.6 \\
		Deep CASA \cite{liu2019divide} & 12.8M & 17.7 & 18.0 \\
		FurcaNeXt \cite{shi2019furcanext} & 51.4M & -- & 18.4 \\
		\hline
		DPRNN-TasNet & \bf{2.6M} & \bf{18.8} & \bf{19.0} \\
		\thline
	\end{tabular}
\end{table}

\subsection{Speech separation and recognition results}

We use a conventional hybrid system for speech recognition. Our recognition system is trained on large-scale single-speaker noisy reverberant speech collected from various sources \cite{YoshiokaEtAl:ms-tr2019}. Table~\ref{tab:asr} compares TCN- and DPRNN-based TasNet models with a 2-ms window (32 samples with 16 kHz sample rate). We can observe that DPRNN-TasNet significantly outperforms TCN-TasNet in SI-SNRi and WER, showing that the speriority of DPRNN even under challenging noisy and reverberant conditions. This further indicates that DPRNN can replace conventional sequential modeling modules across a range of tasks and scenarios.

\begin{table}[!htbp]
	\small
	\centering
	\caption{SI-SNRi and WER results for noisy reverberant separation and recognition task. Window size is set to 32 samples for both models, and the chunk size is set to 100 frames for DPRNN-TasNet. WER is calculated for both separated speakers.}
	\label{tab:asr}
	\begin{tabular}{c|c|c|c}
		\thline
		\thead{Separator\\ network} & \thead{Model \\ size} & \thead{SI-SNRi \\ (dB)} & \thead{WER \\ (\%)} \\
		\hline
        TCN & 5.1M & 7.6 & 28.7 \\
        DPRNN & \bf{2.6M} & \bf{8.4} & \bf{25.9} \\ \hline
		Noise-free reverberant speech & -- & -- & 9.1 \\
		\thline
	\end{tabular}
\end{table}

\section{Conclusion}
\label{sec:conclusion}
In this paper, we proposed dual-path recurrent neural network (DPRNN), a simple yet effective way of organizing any types of RNN layers for modeling an extremely long sequence. DPRNN splits the sequential input into overlapping chunks and performs intra-chunk (local) and inter-chunk (global) processing with two RNNs alternately and iteratively. This design allows the length of each RNN input to be proportional to the square root of the original input length, enabling sublinear processing and alleviating optimization challenges. We also described an application to single-channel time-domain speech separation using time-domain audio separation network (TasNet). By replacing 1-D CNN modules with deep DPRNN and performing sample-level separation in the TasNet framework, a new state-of-the-art performance was obtained on WSJ0-2mix with a 20 times smaller model than the previously reported best system. Experimental results of noisy reverberant speech separation and recognition were also reported, proving DPRNN's effectiveness in challenging acoustic conditions. These results demonstrate the superiority of the proposed approach in various scenarios and tasks.

\bibliographystyle{IEEEbib}
\bibliography{refs}

\end{document}